\documentclass{aastex63}
\usepackage{mathrsfs}
\usepackage{ gensymb }
\received{January 1, 2021}
\revised{January 1, 2021}
\accepted{\today}
\submitjournal{ApJ}

\shorttitle{HXMT observing a solar flare}
\shortauthors{Zhang et al.}
\graphicspath{{./}{hxmtfig/}}

\begin{document}

\title{Non-thermal electron energization during the impulsive phase of an X9.3 flare revealed by Insight-HXMT}

\correspondingauthor{Wei Wang}
\email{wangwei2017@whu.edu.cn}

\author{P. Zhang}
\affiliation{Department of Physics and Technology,Wuhan University Wuhan,430072,China}
\affiliation{WHU-NAOC Joint Center for Astronomy,Wuhan University Wuhan,430072,China}

\author{W. Wang}
\affiliation{Department of Physics and Technology,Wuhan University Wuhan,430072,China}
\affiliation{WHU-NAOC Joint Center for Astronomy,Wuhan University Wuhan,430072,China}

\author{Y. Su}
\affiliation{Purple Mountain Observatory of the Chinese Academy of Sciences, Nanjing, China}

\author{L.M. Song} \affiliation{Institute of High Energy Physics of the Chinese Academy of Sciences, Beijing, China}
\affiliation{University of Chinese Academy of Sciences, Chinese Academy of Sciences, Beijing 100049, China}

\author{C.K. Li}
\affiliation{Institute of High Energy Physics of the Chinese Academy of Sciences, Beijing, China}

\author{D.K. Zhou} \affiliation{Institute of High Energy Physics of the Chinese Academy of Sciences, Beijing, China}
\affiliation{University of Chinese Academy of Sciences, Chinese Academy of Sciences, Beijing 100049, China}

\author{S.N. Zhang} \affiliation{Institute of High Energy Physics of the Chinese Academy of Sciences, Beijing, China}
\affiliation{University of Chinese Academy of Sciences, Chinese Academy of Sciences, Beijing 100049, China}

\author{H. Tian}
\affiliation{The School of Earth and Space Sciences, Peking University, Beijing, China}

\author{S.M. Liu}
\affiliation{Purple Mountain Observatory of the Chinese Academy of Sciences, Nanjing, China}

\author{H.S. Zhao}
\affiliation{Institute of High Energy Physics of the Chinese Academy of Sciences, Beijing, China}
\author{S. Zhang} \affiliation{Institute of High Energy Physics of the Chinese Academy of Sciences, Beijing, China}
\begin{abstract}
The X9.3 flare SOL20170906T11:55 was observed by the CsI detector aboard the first Chinese X-ray observatory Hard X-ray Modulation telescope (Insight-HXMT). By using wavelets method, we report about 22s quasiperiodic pulsations(QPPs) during the impulsive phase. And the spectra from 100 keV to 800 keV showed the evolution with the gamma-ray flux, of a power-law photon index from $\sim 1.8$ before the peak, $\sim 2.0$ around the flare peak, to $\sim 1.8$ again. The gyrosynchrotron microwave spectral analysis reveals a $36.6\pm0.6 \arcsec$ radius gyrosynchrotron source with mean transverse magnetic field around 608.2 Gauss, and the penetrated $\ge$ 10 keV non-thermal electron density is about $10^{6.7} \mathrm{cm}^{-3}$ at peak time. The magnetic field strength followed the evolution of high-frequency radio flux. Further gyrosynchrotron source modeling analysis implies that there exists a quite steady gyrosynchrotron source, the non-thermal electron density and transverse magnetic field evolution are similar to higher-frequency light curves. The temporally spectral analysis reveals that those non-thermal electrons are accelerated by repeated magnetic reconnection, likely from a lower corona source.
\end{abstract}



\section{Introduction} \label{sec:intro}
Super flare releases bulk of energy by various radiation processes in the whole wavebands. From Sep 4 to Sep 10 2017, numbers of big flares occurred, including  tens of M-class and 4 X-class flare \citep{2017RNAAS...1...24S}, also resulted in huge impacts on space environment. Even though SOL2017-09-06T11:55 X9.3 flare is the strongest one in the solar cycle 24, there are limited observational data from ground to space compared with the X8.2 flare on Sep 10, which has been well studied by large numbers of multi-wavelength data and numeric modeling \citep{2020NatAs.tmp..150C}. Furthermore \cite{2019ApJ...877..145L} gave a comprehensive study of gamma ray emission from the X9.3 flare, including the electron-positron 511 keV line, neutron capture line, and showed that the different spectral components are very sensitive to the distribution of accelerated ions. However the broken power law photon spectrum shows a miner difference $\Delta \gamma \leq 0.5$, which could be interpreted as a single power law non-thermal electrons via bremsstrahlung process emitting high energy hard X-ray photons.

Somehow, the electrons above tens of keV also as the very efficient emitters can produce radio emission in the corona, especially for those electrons higher than 30 keV, which spiral in the corona magnetic fields emitting very high frequency radio emission by gyrosynchrotron process. Those electrons are accelerated and injected in magnetohydrodynamic plasma flux tubes, the receptive energy releases would lead to plasma instabilities or magnetohydrodynamic waves modulation \citep{2019NatCo..10.2276C,2020STP.....6a...3K}, in turn modulate the accelerated particles, so as to produce a quasiperiodic pulsations(QPP) both in radio and X-ray bands \citep{2009SSRv..149..119N,2016SSRv..200...75N}. The QPPs exist in solar flares from all wavebands, within a time range from seconds to few minutes\citep{2016ApJ...833..284I}. Recently by timing analysis of the flare SOL20170906T11:55, \cite{2020ApJ...888...53L}  found a $\sim 20$ s QPP in the hard X-rays, $\gamma$-rays and 1.250 GHz radio band. In addition, there exists the more complicated QPPs probed by \cite{2020ApJS..250...31K} using the 22-5000 MHz data, which presented drifting and bi-directional pulsation structures.

It's well known that in the typical flare, energetic electrons are accelerated upwards and downwards during the magnetic re-connection, with a rather short time scale as short as $10^{-5}$ s. Energetic electrons reach the dense bottom of corona or chromosphere, losing their energy through collisions, and produce hard X-ray impulses \citep{2009A&A...504.1057S}. Especially those mildly relativistic energetic electrons, come to be very effective emitter of hard X-ray photons in the flare source. It also should be noticed that in the corona the electron gyrofrequencies $f[\mathrm{Hz}] \sim 2.8\times 10^6 B[\mathrm{G}]$, which will yield order of GHz radio emissions due to few hundreds gauss magnetic filed.
So the simultaneous fine resolution of spatial and timing observations of hard X-ray and radio observations with gyrosynchrotron process will reveal the local plasma source properties, magnetic field configuration and energetic electrons distribution in the local source region \citep{2020Sci...367..278F,2020NatAs.tmp..150C,2020arXiv200615014Z}. This is particularly important because the coronal magnetic field is very difficult to measure, though recently \cite{2020ScChE..63.2357Y}  and \cite{2020Sci...369..694Y} have measured the magnetic field in the off-limb corona by combing spatial distribution of the plasma density and the phase speed of the prevailing transverse magnetohydrodynamic waves, based on near-infrared imaging spectroscopy observations.

In our study, we will present the timing and spectral analysis of the X9.3 flare impulsive phase from the Insight-HXMT observation and RSTN (Radio Solar Telescope Network, San vito) data. We shows the hard X-ray light-curves observed from HXMT/CsI detector, and its quasiperiodic pulsations (QPPs) by wavelet analysis. We also apply the same method to the RSTN data, presenting an interesting time delay for hard X-ray for the same QPP. We selected several time intervals within hard X-ray impulsive phase to perform the spectral analysis, and found a very hard power law distribution of photon flux in the rang 100-800 keV, but with a variability during the impulsive phase. The radio spectra on the other hand tend to be very complex, thus we only model the peak time with a homogeneous cylinder hot plasma tube with simplified magnetic configuration by gyrosynchrotron process, and the results indicated that during the impulsive phase, the non-thermal electrons density and magnetic field is rather steady. In Section 2, the observations of Insight-HXMT and RSTN are introduced. We showed the model used to estimate the corona plasma properties in Section 3. The detailed spectral analysis combined with Insight-HXMT and RSTN are presented in Section 4. The magnetic field and electron evolution of the GS source are studied in Section 5. Finally the conclusion and summary are given in Section 6.

\section{observations} \label{sec:obs}

Insight-HXMT\footnote{http://hxmten.ihep.ac.cn} is a dedicated hard X-ray telescope designed for all-sky surveys of pulsars, neutron stars and black holes in X-ray binaries from 1-250 keV \citep{2020SCPMA..63x9502Z} with High energy telescope (HE), Medium Energy telescope (ME) and Low Energy telescope (LE), also have the capacity of monitoring electromagnetic counterparts of gravitational wave sources from 0.1-3 MeV with the HE. The HE telescope constructs with 18 cylindrical NaI/CsI detectors, NaI detector was designed for nominal filed of view observations from 20-250 keV, CsI detector was designed for gamma rays from 80-800 keV(Normal-Gain mode) and 200-3000 keV(Low-Gain mode) respectively \citep{2020SCPMA..6349503L}. Due to the crystal thickness limit and small filed of view of NaI detector, GRBs or Solar flares are very hard to detected by NaI detector. However the CsI detector has a very large effective area which is sensitive to high energy gamma rays. In-orbit calibration by combining Crab pulse radiation and on-ground simulation show that the CsI detector performs in a very stable state, and joint cross in-orbit calibrations with Fermi/GBM,Swift/BAT and KONUS-WIND proved that the CsI detector has been well calibrated \citep{2020JHEAp..27....1L}, and tend to be a very promising large effective area detector for $> 100$ keV gamma ray emission source.

\begin{figure}[ht!]
   \epsscale{0.85}
\plotone{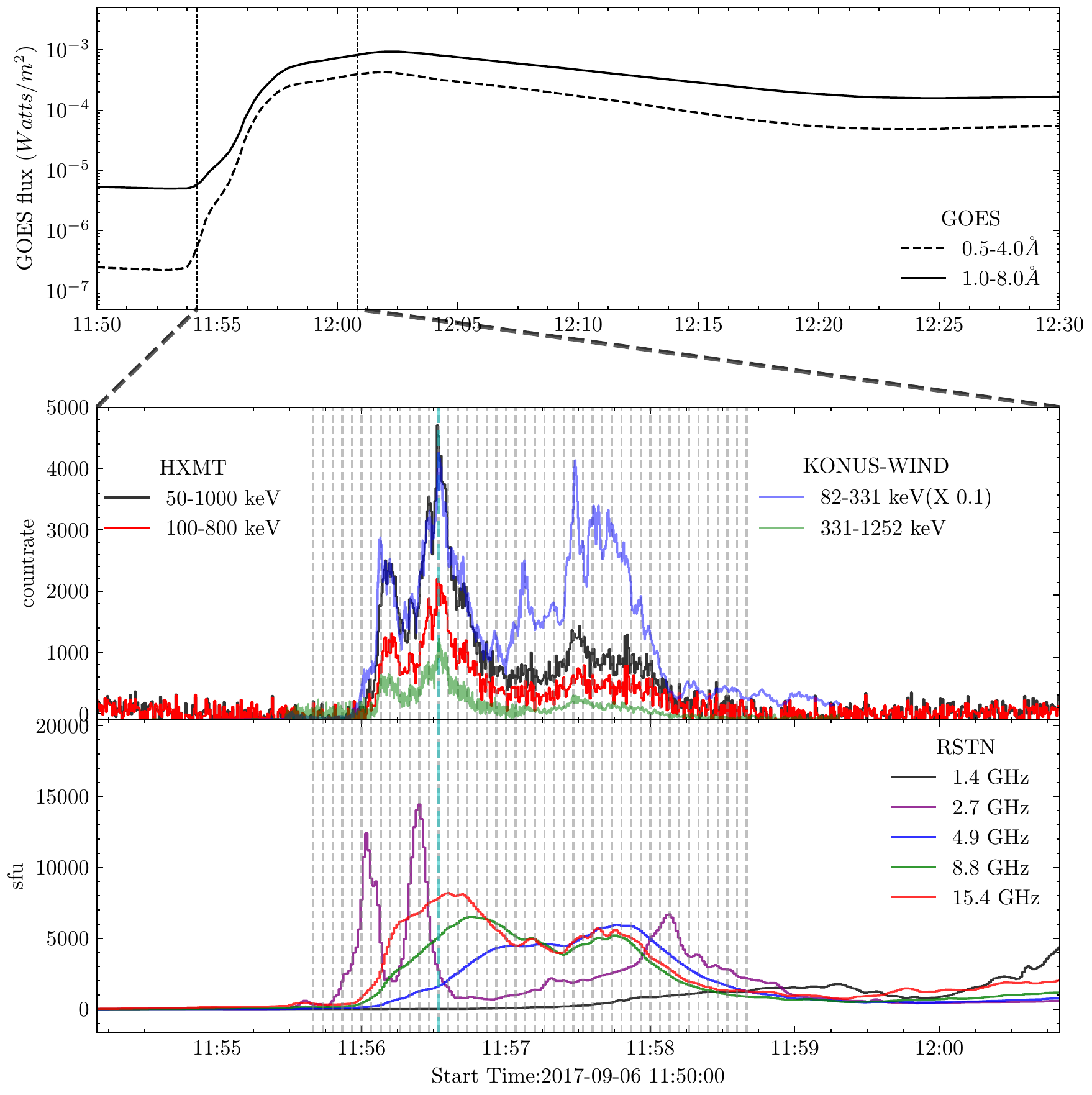}
\caption{The multi-wavelength light curves of SOL2017-09-06T11:55: (upper panel) the GOES soft X-ray light curves, the dashed line is $0.5-4.0 \AA$, solid line is $1.0-8.0 \AA$; (middle panel) the Insight-HXMT hard X-ray light curves from 50-1000 keV, 100-800 keV, and KONUS-Wind light curves from 82-331 keV (G2 band) and 331-1252 keV(G3 band); (the last bottom panel) radio light-curves in 1.4GHz, 2.7GHz, 4.9GHz, 8.8GHz and 15.4GHz from RSTN observations. We marked the spectral interval with gray dashed vertical lines. The HXMT light-curve main peak was also marked with cyan dashed line at 11:56:32 UT. \label{fig:lc}}
\end{figure}

The active region 12673 became very productive during 4 Sep to 10 Sep 2017, and the SOL2017-09-06T11:55 X9.3 flare came to be the strongest flare in the last solar cycle.  In figure \ref{fig:lc}, we could see that light-curves from soft X-ray , hard X-ray and radio bands showed a very significant enhancement during the flare. AS RHESSI (Reuven Ramaty High-energy Solar Spectroscopic Imager) and Fermi (The Fermi Gamma-Ray Space Telescope) entered night shade during the main phase of the X9.3 flare, we got very limited gamma ray emission observations for this event.

The HXMT CsI detector and KONUS-WIND (KW) NaI detector both detected the flare peak phase, and HXMT performed a stable observation with Low-Gain mode during the main hard X-ray impulsive phase (see the light curves in the figure \ref{fig:lc}, the hard X-ray light curves have already subtracted background). In the soft X-ray band measured by GOES(Geostationary Operational Environment Satellite), the thermal bremsstrahlung emission started growing at around 11:53:45 UT. As RHESSI and FERMI both entered night shade, we lacked observations of this impulsive phase with these two instruments. And the HXMT observations show a good agreement with KONUS-WIND G2 band(82-331 keV) and G3 band(331-1252 keV) observation during the hard X-ray impulsive phase. Because the HXMT CsI detector has higher threshold for incoming gamma ray photons than KONUS-WIND, that only photons energy bigger than 200 keV could fully penetrate and deposit into the detector, considering the photons could be also scattered by satellite itself which means the incoming photon energy is bigger than the deposit photons. Moreover the counts rate difference may be due to the effective area differences between HXMT and KONUS-WIND \citep{2020JHEAp..27....1L}. KONUS-WIND has 80-160 $\rm cm^2$ effective area for each detector \citep{1995SSRv...71..265A}, but HXMT/CsI has got less than $10^2$ $\rm cm^2$ below 100 keV and raises to more than $10^3$  $\rm cm^2$ above 200 keV. During the flare , we could also get the sun position is (RA $\sim$ 165.1$\degree$, Dec $\sim$ 6.4$\degree$), and the HXMT detector position is (RA $\sim$ 233.8$\degree$, Dec $\sim$ -57.2$\degree$), which indicate the flare photons comes from pitch angle $\sim$ -63.6 $\degree$ and incident from the back side. Synthesis of these views of the difference between HXMT/CsI and KW, we could see the count rate discrepancy of these two instruments. At low energy band the KW G1 and G2 observed photons up to $10^4$ keV, which is much higher than HXMT. It should be noted that in the figure \ref{fig:lc}, we has multiplied 0.1 for G2 band light curves for a better comparison with HXMT observations.

In the last panel of figure \ref{fig:lc}, we showed the centimeter microwave measurements from RSTN (Radio Solar Telescope Network,San vito) at 1.4,2.7,4.9,8.8 and 15.4 GHz \citep{1981BAAS...13Q.553G,2018SoPh..293...50T}.  The radio emission at 1.4GHz has a rather lower flux density than those at higher frequencies during the X9.3 flare, the pattern of 2.7 GHz flux showed several impulsive peaks during the impulsive phase. Radio emissions at 4.9 GHz, 8.8 GHz and 15.4GHz started brightening from 11:56 UT, showing relatively similar trends with hard X-ray observations which indicate strong gyrosynchrotron processes. And the few seconds time delay between hard X-ray and GHz radio peaks may comes from a consequent of energetic electron acceleration time scale difference in the flare region.  The dashed lines in figure \ref{fig:lc} are the ROI(regions of interest) time intervals for further hard X-ray and radio spectral analyses with 4 seconds integration.

\begin{figure}[ht!]
\gridline{\fig{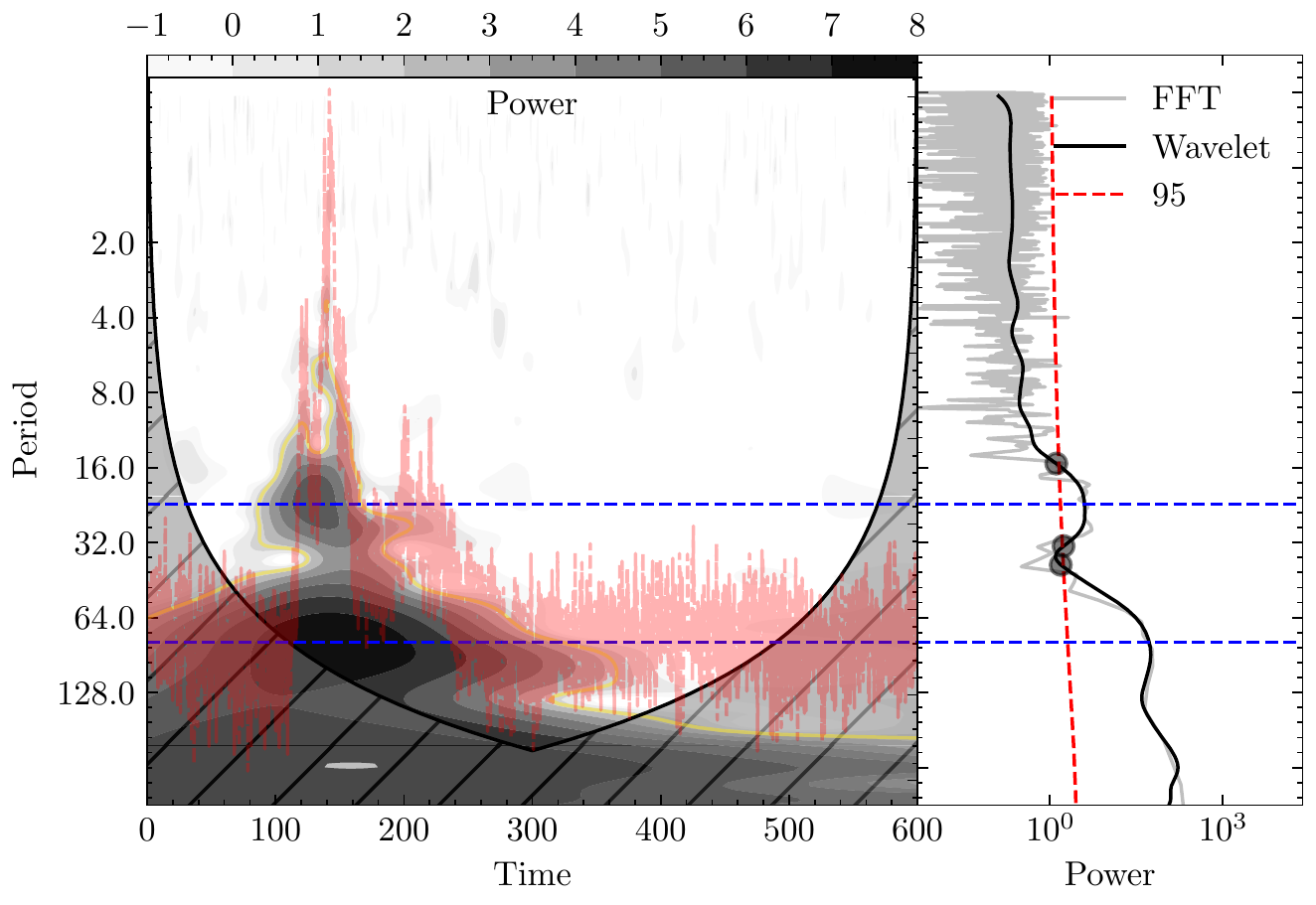}{0.5\textwidth}{(a)}
          \fig{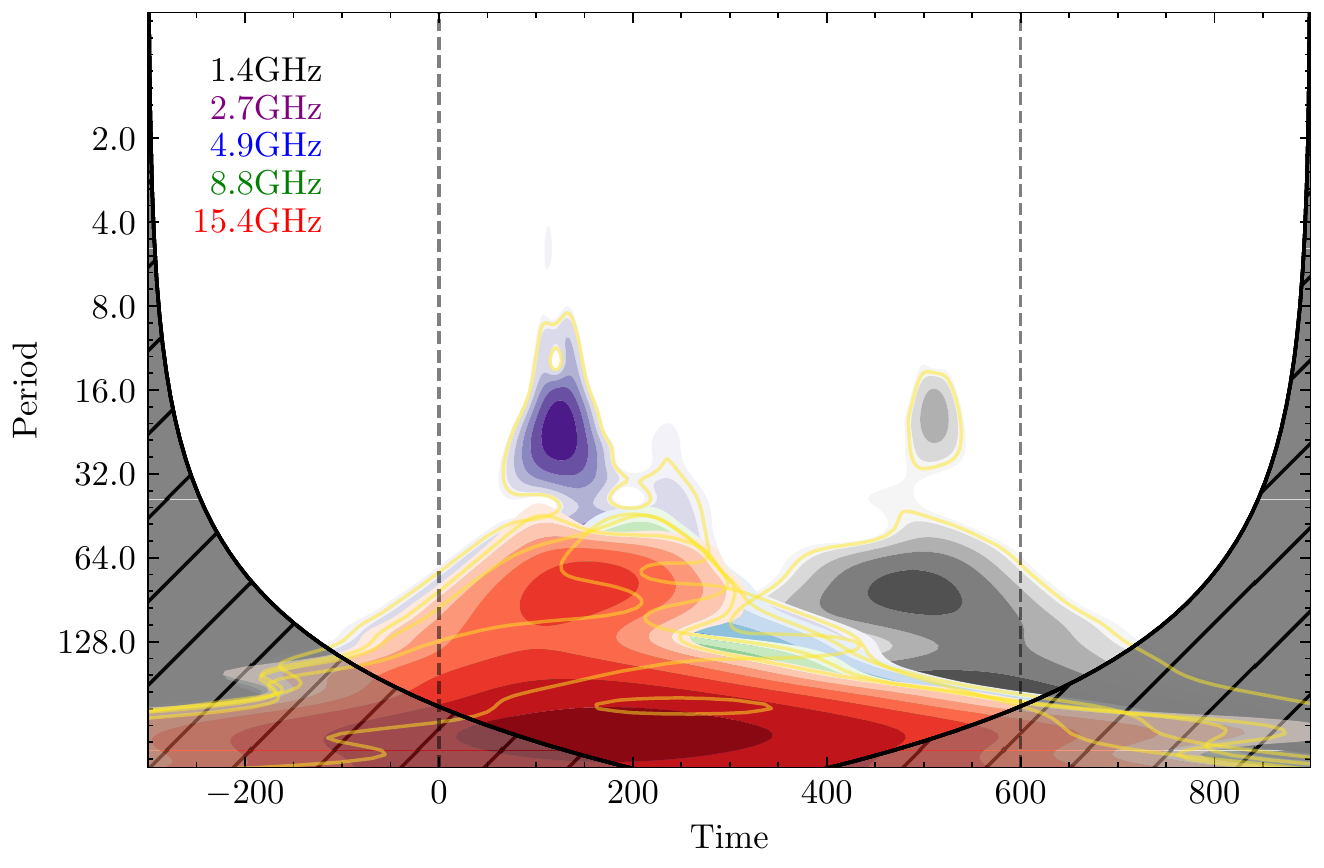}{0.5\textwidth}{(b)}
          }
\caption{The wavelet analysis of the X9.3 light curves from HXMT and RSTN observations, both setting 11:54:10 UT as time zero. (a) Wavelet results of the HXMT light curve in 100-800 keV range, left panel is the wavelet power contour along time axis with cone of influence (COI) curve over plot, the normalized light curve also over plot; the right panel showed the global power spectrum from FFT (gray line), wavelet(black line) and the white noise level(red dash), FFT power spectrum re-normalized to match wavelet global power spectrum ; (b) using the same wavelet method applied to RSTN radio data, the wavelet power spectra are over-plotted for 1.4 GHz , 2.7 GHz , 4.9 GHz , 8.8 GHz and 15.4 GHz with black, purple, blue, green and red contours respectively, the two dashed black vertical line marks the HXMT light curve time range.} \label{fig:wave}
\end{figure}

 We also used wavelet method to investigate the quasiperiodic pulsation for the flare light curves as shown in figure \ref{fig:lc}. The wavelet transform technique is based on \cite{1998BAMS...79...61T}, which provides a distinguished method for intrinsic timing pattern detection. It has already been used in detection of quasiperiodic pulsation in many respects, also including solar flares. For the X9.3 flare, it was found the modulation periods 24-30s from 11:57 UT to 11:58 UT in hard X-rays, and with a slightly short period 20s from 11:55:30 UT to 11:57 UT, but in the radio spectrum 22-5000 MHz, there exists a significant broad band QPP range from few seconds to 50s during the main phase of the X9.3 flare (11:55 UT to 12:07 UT) \citep{2020ApJ...888...53L,2020ApJS..250...31K}.

 In this study, we use the Morelet function as the mother function of wavelet transform, and the time resolution is 0.5 s for hard X-ray data and 1.0 s for the RSTN data respectively. Here we also hypothesise that the background noise from  hard X-ray, $\gamma$-ray and high frequency radio band ($\ge$ GHz) is white noise, so in the wavelet analysis we have taken a white noise spectrum for significance calculation, and the confidence level was set to 95$\%$. Figure \ref{fig:wave} (a) shows the 100-800 keV HXMT light curve wavelet transform, a very distinct periodic pattern present from 11:57 UT to 11:59 UT  with QPPs at period of $\sim 22^{+11}_{-7} \rm s$, and the period of $80\pm 41 \rm s$, which is in agreement with QPPs detected by KONUS-WIND data \citep{2020ApJ...888...53L}. The QPP at $\sim 80$ s was almost located inside the COI which could be not distinguished from edge data effect.  The RSTN radio light curves wavelet transform presented in the figure \ref{fig:wave} (b). The lower frequency (1.4 GHz) wavelet power spectrum indicated various QPPs and drifts which were confined by \cite{2020ApJS..250...31K}. The 2.7 GHz curve has a similar QPP period at $\sim 20$ s with the 100-800 keV curve but with few seconds ahead in time. However, higher frequencies do not show any obvious short time scale QPPs. As shown in figure \ref{fig:wave} (b)  the longer time wavelet shows a QPP period at $\sim 80 s$, which is also corresponding with hard X-ray, suggesting that the  gyrosynchrotron source tends to be rather steady in a few minutes during the impulsive phase.

\section{Plasma configuration and modeling}\label{sec:model}

Thermal bremsstrahlung emission from hot corona plasma contributes to the gradual phase of flares in general, and the intensive impulsive emission mainly comes from gyrosynchrotron emission process\citep{2018A&A...615A..48Z}. To quantify the non-thermal energy releasing process from the X9.3 flare, especially the non-thermal energetic electrons which were the primary contributor of the impulsive peak phase for hard X-ray emission, we have to combine the limited hard x-ray observations and radio observation to build an evolutionary gyrosynchrotron source model. However, it's true that the magnetic reconfiguration which powers the flare energy release always indicates the complex topology and timing evolution corona plasma and magnetic filed configuration\citep{2019ApJ...880L..29A}. In the X9.3 flare, as shown in the figure \ref{fig:lc}, we've got radio flux up to $10^4$ $\rm sfu$, hence here we assume such a simple model, that a homogeneous column source with line-sight length $L$, with a thermal electron density $n_e$ estimated using $EM=\int{n_{e}^2dV}$, and the energetic electrons accelerated  lose their energy in the dense hot plasma column tube, spiral in the corona magnetic filed emitting higher frequency radio radiation through gyrosynchrotron process.

\begin{figure}[hbt!]
\epsscale{0.85}
\plotone{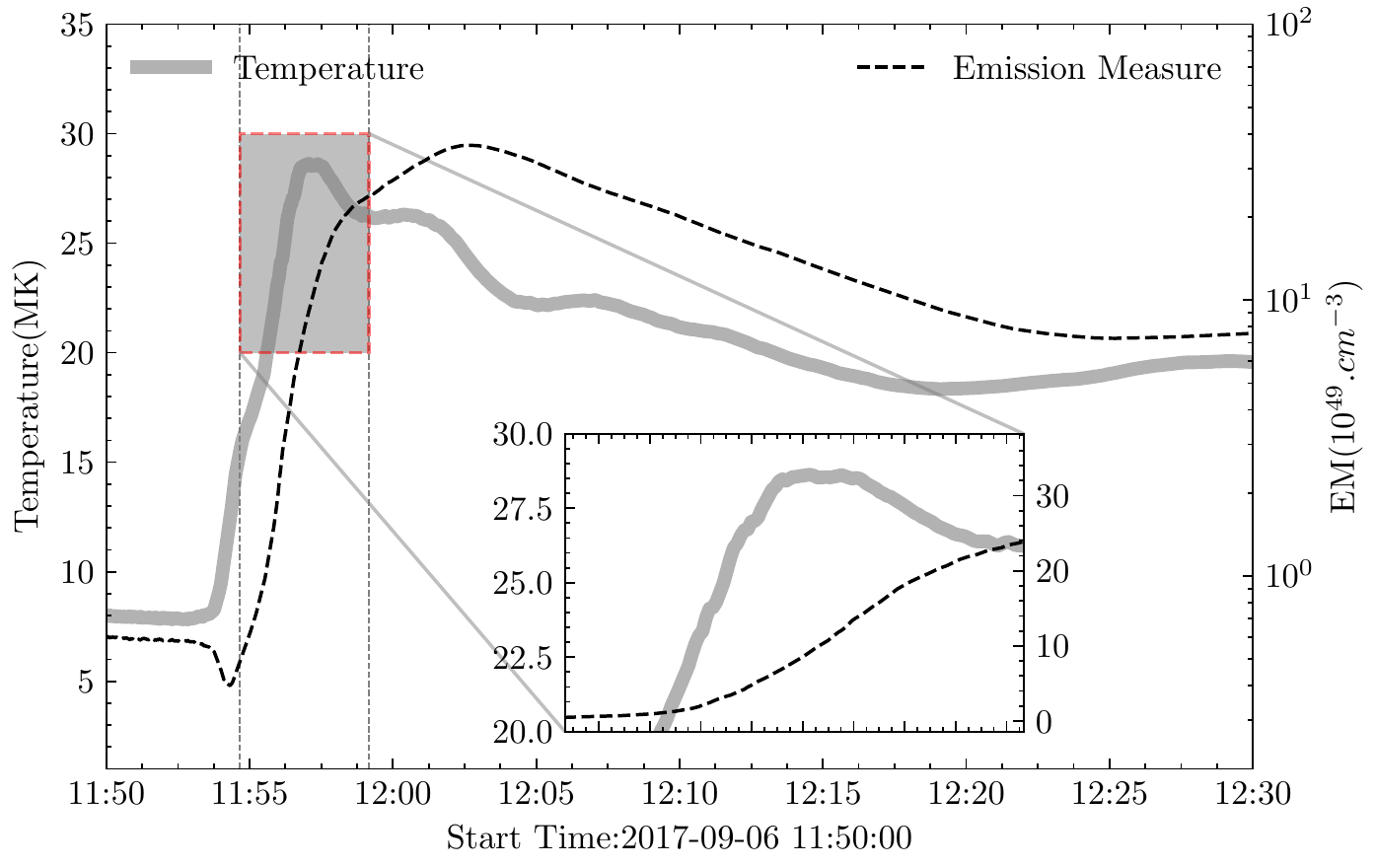}
    \caption{The figure shows the X9.3 flare active region plasma evolution, temperature and emission measure inferred from GOES data, zoom figure inside is the ROI time intervals.}
    \label{fig:emtau}
\end{figure}

In the figure \ref{fig:emtau}, we derived the corona plasma emission measure and mean temperature from GOES \citep{1994SoPh..154..275G}. The GOES measured soft X-rays at $0.5-4.0 \AA$ and $1.0-8.0 \AA$, which are sensitive to a few MK(Million Kelvin) to tens of MK plasma. We could see the onset of the X9.3 flare from GOES observations starting from 11:53:45 UT, corona temperature raised rapidly to more than 20 MK during the peak phase and reached its peak around 11:57$\sim$11:58 UT, when plasma emission measure boosted concurrently and reached its peaks around 12:03:45 UT. Such rapid corona heating leads to the super-thermal electrons trapped in the plasmoid or loops which also could be a possible interpretation of drift QPPs in the lower radio frequency range \citep{2000A&A...360..715K}.

For given homogeneous flare source, the optical depth $\tau=0.2\frac{EM}{T_{\rm e}^{1.5}f^2A_{\rm source}}$, is inversely proportional to the square of emission frequency(f), at lower frequency is optical thick ($\tau \gg 1$) and for higher frequency is optical thin($\tau \ll 1$). The $\frac{1}{T_{\rm e}^{1.5}}$ indicate cool dense plasma is very efficient in producing thermal bremsstrahlung emission, but the GOES soft X-ray band is sensitive to $\ge$ 10 MK plasma, so we often see the inferred thermal bremsstrahlung radio emission is few times lower than the observation flux if we only use GOES data. Moreover the thermal bremsstrahlung radio flux is proportional to the optical depth $\tau$, so during the plasma heating phase, higher frequency (15.4 GHz) usually became optical thin, and when hot plasma start to cooling down become optical thick to lower frequencies where the radio emission mainly comes from coherent mechanism (plasma emission or cyclotron emission)\citep{1985ARA&A..23..169D,2011SSRv..159..225W}. But the inferred thermal bremsstrahlung flux in RSTN radio band is of order 50-180 sfu during the X9.3 flare, about 100 times lower than observation flux, imply the thermal emission for higher frequencies radio emission only at background level, and basically gyrosynchrotron emission dominate the main phase for the centimeter radio observation during the X9.3 flare.

\begin{figure}[hbt!]
\epsscale{0.75}
\plotone{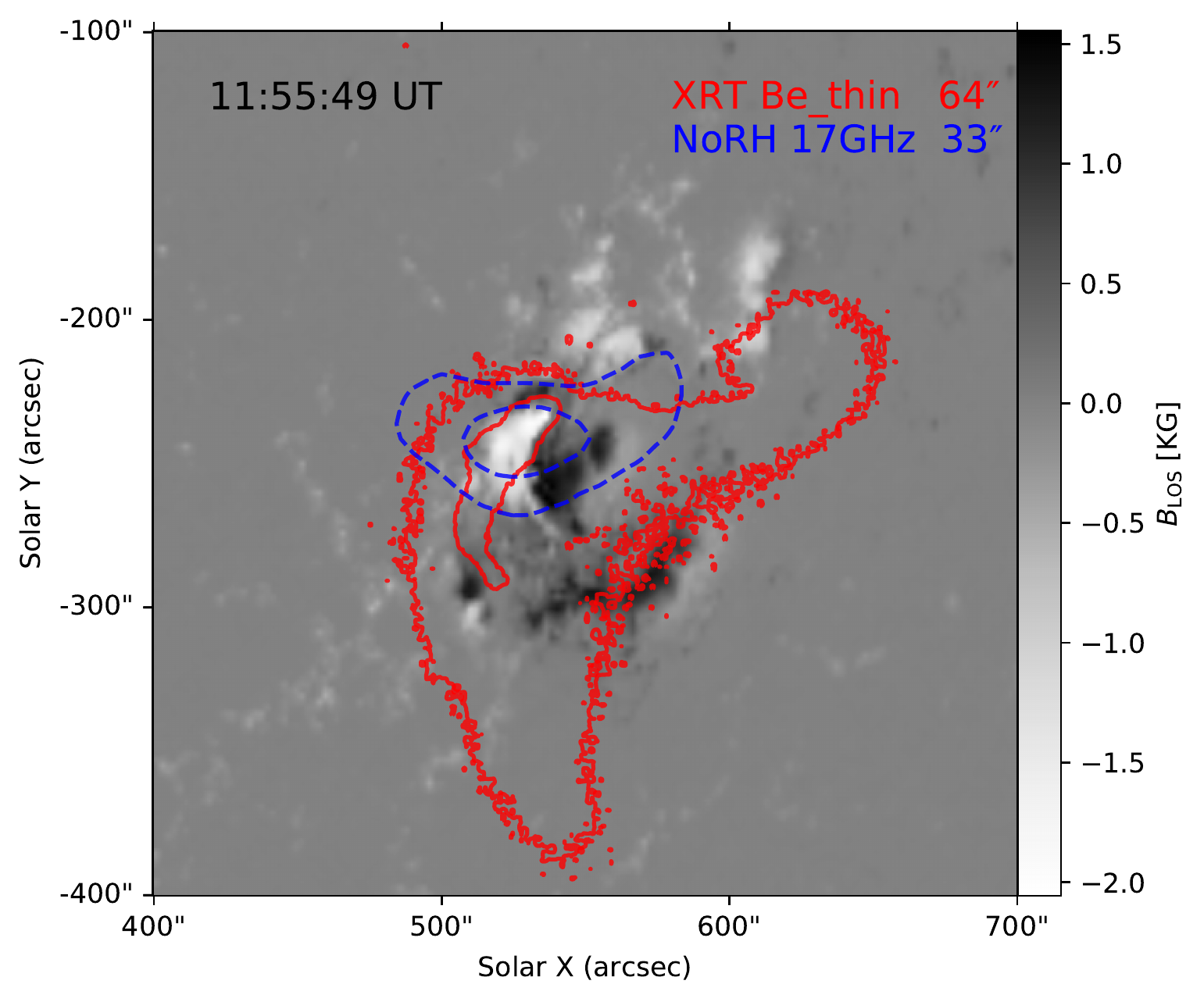}
    \caption{The figure shows the X9.3 flare active region at 11:55:49 UT, background image is the HMI LOS magnetic, overplot contours from Hinode/XRT Be$\_$thin observation and NoRH 17 GHz observations at 23:36 UT with solar rotation correction evolution. The contour levels are $1/{e^2}$ and $50\%$, the beam diameter for soft X-ray is 64$\arcsec$, and for 17GHz is 33$\arcsec$ .}
    \label{fig:active}
\end{figure}

 Even though \cite{2019ApJ...880L..29A} has presented the existence of a long live gyroresonant(GR) emission source during Sep 6, we lack radio image observation during the X9.3 flare peak phase (only the Metsähovi Radio Observatory operated by the Aalto University in Finland got 37 GHz Radioheliograph maps, see figure \ref{fig:mromap} in the appendix, but Metsähovi Radio Observatory observation has rather low  spatial and temporal resolution solar maps for source size estimation). Thus we have taken the NoRH 17 GHz maps at 23:36 UT (peek time of a M class flare at Sep 6) to estimate lower limit of the radio source size. The source beam size in this study was estimated by measuring the beam diameter (the source $\ge \frac{1}{e^2}$ of max source intensity, then the equivalent beam diameter$R_{\rm source}=\sqrt{A_{\rm source}/\pi}$ ), as shown in the figure \ref{fig:active}, thus the inferred 17 GHz source is about 33$\arcsec$ after applying solar rotation from 23:36:00 UT to 11:55:49 UT, which was perceived as a gyrosynchrotron(GS) source lower limit. However the Hinode/XRT Be$\_$thin observations presented a soft X-ray source with about 64$\arcsec$ beam diameter, $\ge$ MK hot plasma source extended along the light bridge at the beginning of the X9.3 flare. We also estimated the $50\%$ contour level source beam size of 17$\arcsec$ and 15$\arcsec$ for sort X-ray and radio respectively, both well correlated with photosphere line of sight magnetic filed, indicating that the main GS source is located near the light bridge at the foot point of the flare.



\section{Spectral analysis}\label{sec:spect}
To model the GS source in the X9.3 flare, we have taken this hypothesis that the injected beaming energetic electrons configuration is power law distribution with low energy cutoff and high energy cutoff.
For a single power law hard X-ray spectrum , the energetic electrons penetrate into the dense corona,  lose all their energy in the thick target emitting hard X-rays\citep{2011SSRv..159..225W}. The electron distribution could be given by:

 \begin{equation}
\frac{d^{2} \mathscr{N}(E)}{d E d V}=3.04 \times 10^{24} \frac{A_{0} b(\gamma)}{E_{0, \mathrm{keV}}^{1.5} A_{X}}\left(\frac{E}{E_{0}}\right)^{-(\gamma+1.5)} \text { electrons } \mathrm{cm}^{-3} \mathrm{keV}^{-1}.
\label{eq:ele}
\end{equation}

In the equation \ref{eq:ele}, $A_0$ is the photon spectrum normalization factor at $E_{0}$, and $A_X$ is the hard X-ray source area, $b(\gamma)$ could be calculated in given $\gamma$, then the electron distribution has a power law index of $\delta = \gamma+1.5$. Since the low energy electrons lose energy faster and depleted faster than high energy electrons, thus the photon spectrum index is softer than the injection energetic electron spectrum. Then from HXMT observations, we could get the injected energetic electrons distribution.

\begin{figure}[hbt!]
\centering
\gridline{
          \fig{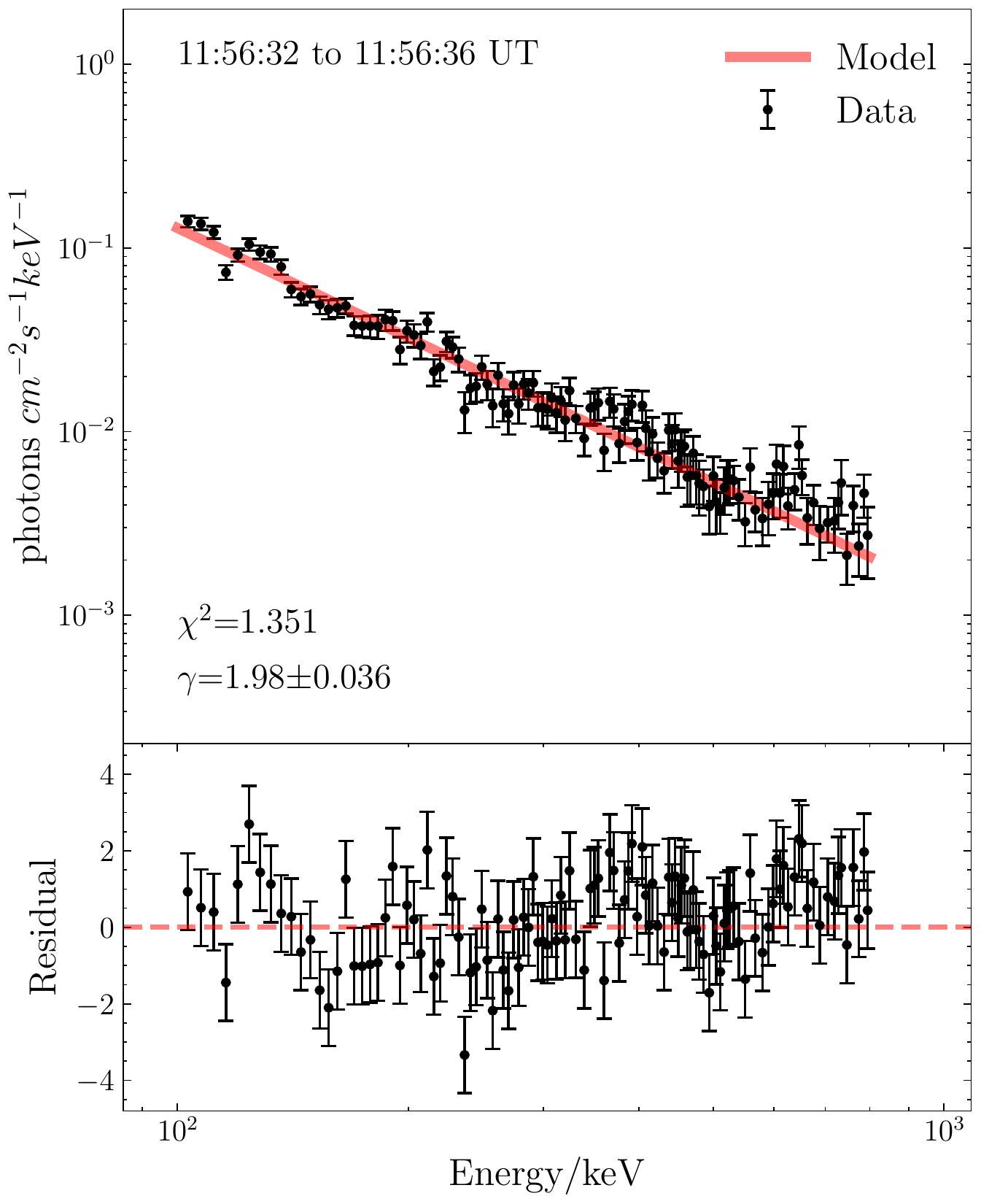}{0.5\textwidth}{(a)}
          \fig{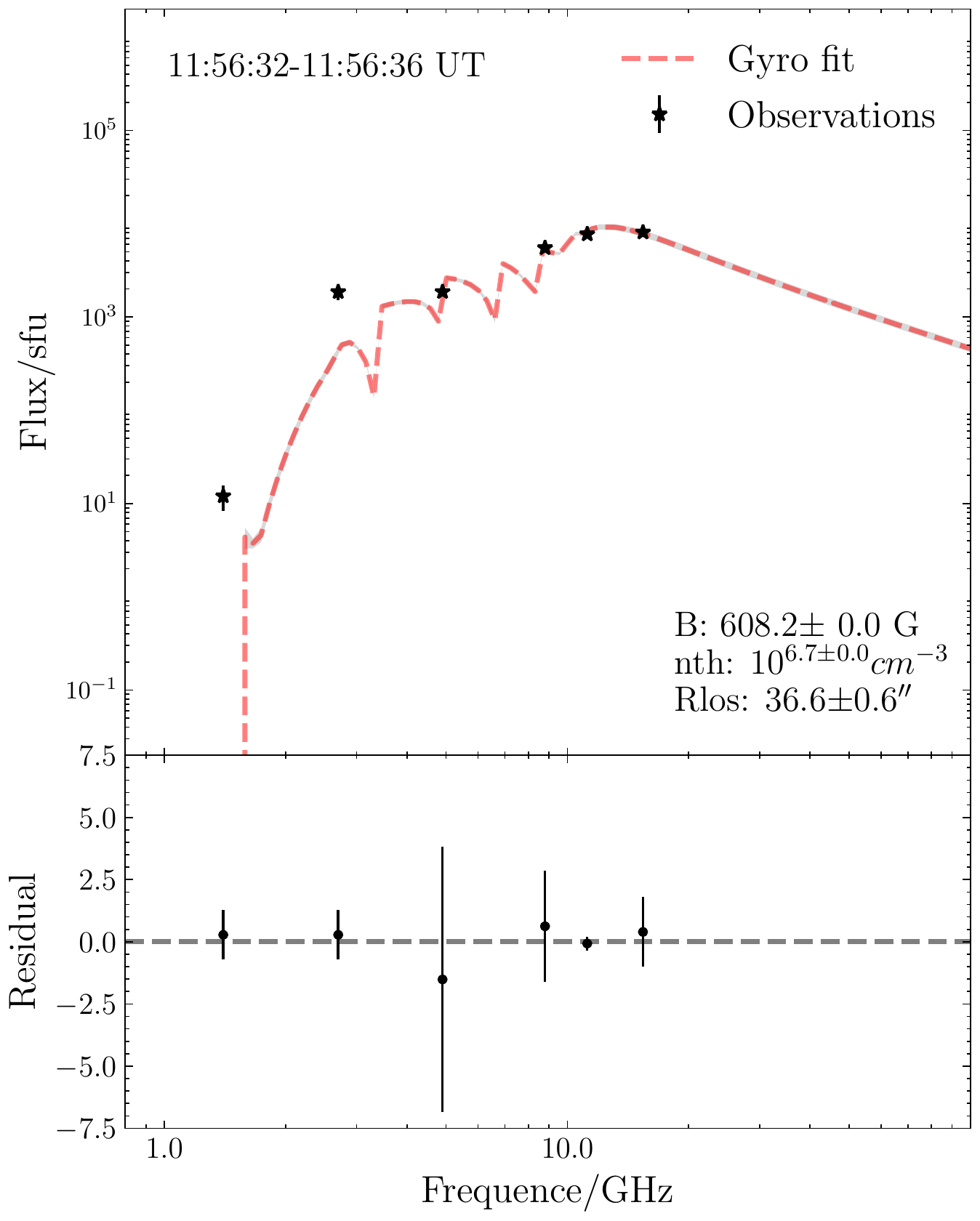}{0.5\textwidth}{(b)}
          }
    \caption{The figure shows the hard X-ray and radio spectrum fitting results of the X9.3 flare from HXMT and RSTN observations. (a) the hard X-ray fitting result at the main hard X-ray peak UT11:57:41 to 11:57:45 ; (b) The radio gyrosynchrotron spectral fitting result at the same time interval. }
    \label{fig:hxmtspec}
\end{figure}

However during the X9.3 flare impulsive phase, we have no hard X-ray spatial information, which makes it difficult to calculate precise energetic electrons flux. Thus we at first derived the hard X-ray spectrum from HXMT observations within energy range 100-800 keV by the HXMT GRB analysis software \footnote{http://hxmtweb.ihep.ac.cn/documents/497.jhtml} and XSPEC\citep{1996ASPC..101...17A} \footnote{https://heasarc.gsfc.nasa.gov/xanadu/xspec/} software from HEASOFT, a single power law model applied in the spectral fitting, as shown in the figure \ref{fig:hxmtspec} (a). The hard X-ray spectrum at the main peak of HXMT light curves at 11:56:32 UT to 11:56:36 UT can be well fitted with a single power law with a $\chi^2 =1.351$, and the photon index $\gamma$ is about $1.98\pm0.036$. It should be noted that we used a statistical error in the fitting. Compared with the spectral fitting results based on KONUS-WIND data from \cite{2019ApJ...877..145L}, the HXMT photon index $>$100 keV is slightly harder. The KONUS-WIND spectra fitting added the nuclear line components above 500 keV plus the power-law component. In addition, the KONUS-WIND data only had spectral data before 11:57:17 UT, but HXMT catched the full phase so we could make a continuous spectral analysis within the ROI time intervals as shown in the section \ref{sec:obs}.

Hence We considered a cylinder gyrosynchrotron radiation source that evolve over time, and performed the fast numerical gyrosynchrotron fitting for RSTN radio data. The gyrosynchrotron fitting already included the contributions of gyrosynchrotron, electron-ion collisions free-free emission and electron-neutral hydrogen atoms free-free emission \citep{2010ApJ...721.1127F}. As the active region locates on the solar disk, we assume a 90$\degree$ angle between the transverse magnetic field and the line-of-sight beam, with dynamic background plasma estimated from GOES observations (the plasma temperature $T_{\rm goes}$ and emission measure $EM_{\rm goes}$), and beaming non-thermal electron distribution from the HXMT observation.

The RSTN microwave observations provided a continuous radio spectrum throughout the ROI time intervals during the main phase of the X9.3 flare, and we also used the 11.2 GHz data from Metsähovi Radio Observatory 1.8-metre radio telescope for better parameter constraint. And we assume the cylinder gyrosynchrotron radiation source line-sight column depth L is 10 $\arcsec$ as typical flare radiation source column depth setting \citep{2020NatAs.tmp..150C,2013A&ARv..21...58K}, and the source area radius $R_{\rm source}$ as free parameter, the radio emission source beam diameter size lower limit from later NoRH 17 GHz image  at 23:36UT of a M class flare is about 33 $\arcsec$ (see  figure \ref{fig:active}), and upper limit about 64 $\arcsec$ constrained by the Hinode soft X-ray Be$\_$thin observation.

Then we inferred the thermal electron density $n_{\rm e,th}=\sqrt{\frac{EM_{\rm goes}}{L\pi*R_{\rm source}^2}}$, as well as the plasma temperature $T_{\rm goes}$ from GOES data. It should be also noticed that the EM and temperature derived from GOES data are lower than SDO/AIA(cool plasma) estimation but higher than RHESSI estimation (hot plasma). In this study we have taken a factor of $10^{-0.6}$ and $10^{0.3}$ for a moderate background plasma source configuration suggested by \cite{2014SoPh..289.2547R}. Moreover the penetrated non-thermal electron beams were set with a power law distribution index $\delta=\gamma_{\mathrm{HXMT}}+1.5$, the optically thin side of radio spectra which refer to high-frequencies were largely dictated by the energetic electron distribution($\delta$). And for electron beam energy cutoff, the statistical study by \cite{2019SoPh..294..105A} showed lower energy cutoff $E_{e,min}$ from few tens keV to hundreds keV for non-thermal electron beams in big flares, hence we fix the low energy cutoff for energetic electrons at 10 keV and a high energy cutoff at $E_{e,max}\sim $10 MeV suggested by \cite{2020NatAs.tmp..150C}. For a comprehensively energetic electron beam configuration, the initial non-thermal electron density $n_{e,nth}$ is set as a free parameter.

It's also well known that the peak frequency of gyrosynchrotron radio spectra is sensitive to both the number density of energetic non-thermal electrons $n_{e,nth}$ and the magnetic field strength B \citep{2010ApJ...721.1127F}. So we set the corona transverse magnetic field B as a free parameter in our fitting. Corona magnetic field is very difficult to measure directly. Recently \cite{2020Sci...369..694Y} and \cite{2020NatAs.tmp..150C} reported corona magnetic field results inferred from magnetohydrodynamic waves and gyrosynchrotron respectively, especially the spatial and temporal resolved corona magnetic field from EOVSA radio data for the Sep 10 X8.2 flare indicated the mean magnetic field for GHz radio source is about few hundred Gauss. Furthermore the radio emission of the X9.3 flare extended to sub-millimeter range\citep{2018SpWea..16.1261G}, reached $10^3$ sfu at 212 GHz and 405 GHz, revealing the strong GS source. \cite{2013A&ARv..21...58K} have proposed a thermal gyrosynchrotron plus non-thermal gyrosynchrotron model, and tried to explain such strong GS emission in similar X class flares, but they constrained a 5200 Gauss magnetic filed, beyond photosphere magnetic limit for the typical active region, which is not reasonable. NLFFF reconstruction of the SDO/HMI magnetogram also showed a long live few kiloGauss(KG) magnetic field present in the base corona \citep{2018RNAAS...2....8W}. And as the magnetic field decreased along the height above the base corona \citep{2020Sci...367..278F}, they suggested the mean corona magnetic field in gyrosynchrotron numerical fitting for the X9.3 flare could be few hundred Gauss.

Even though both \cite{2020NatAs.tmp..150C} and \cite{2020ApJ...904...94S} suggested MCMC algorithm would perform the best error estimation, we have noted that RSTN and MRO radio observations have very limit data points during the X9.3 flare for proper error estimation, thus we choose the fitting algorithm "Nelder-Mead" which integrates in the lmfit \footnote{https://lmfit.github.io/}package as the best algorithm in our fitting after trying various algorithms. Due to the limited radio data, and for the optically thick side of the spectra, thermal plasma free-free absorption and Razin suppression became increasingly important, which leads to very complex spectrum evolution at lower frequencies, and may lead to rather big uncertainties. The 1.4 GHz and 2.7 GHz data showed complicated behaviour during the impulsive phase (see figure \ref{fig:lc}), hence we estimated the error and $\chi^2$ by minimize the residuals only using higher frequencies data during the impulsive phase. The peak time spectral fitting was shown in the figure \ref{fig:hxmtspec} b, and the higher frequencies data have been well fitted in the error space which were marked with the red dot line. At 11:56:32 UT to 11:56:36 UT, when the HXMT main peak time interval occurred, we got a radius about $36.6\pm 0.6 \arcsec$, and a $10 \arcsec$ depth source, which is consistent with the NoRH source estimation. And the local mean transverse magnetic field B was estimated at $\sim$ 608.2 Gauss. Moreover the penetrated non-thermal power-law distribution electron density is about $10^{6.7} cm^{-3}$ for $E_{e}$ from 10 keV to 10 MeV.

\section{Magnetic field and non-thermal electron evolution of GS source}
\begin{figure}[hbt!]
\centering
\epsscale{0.85}
\plotone{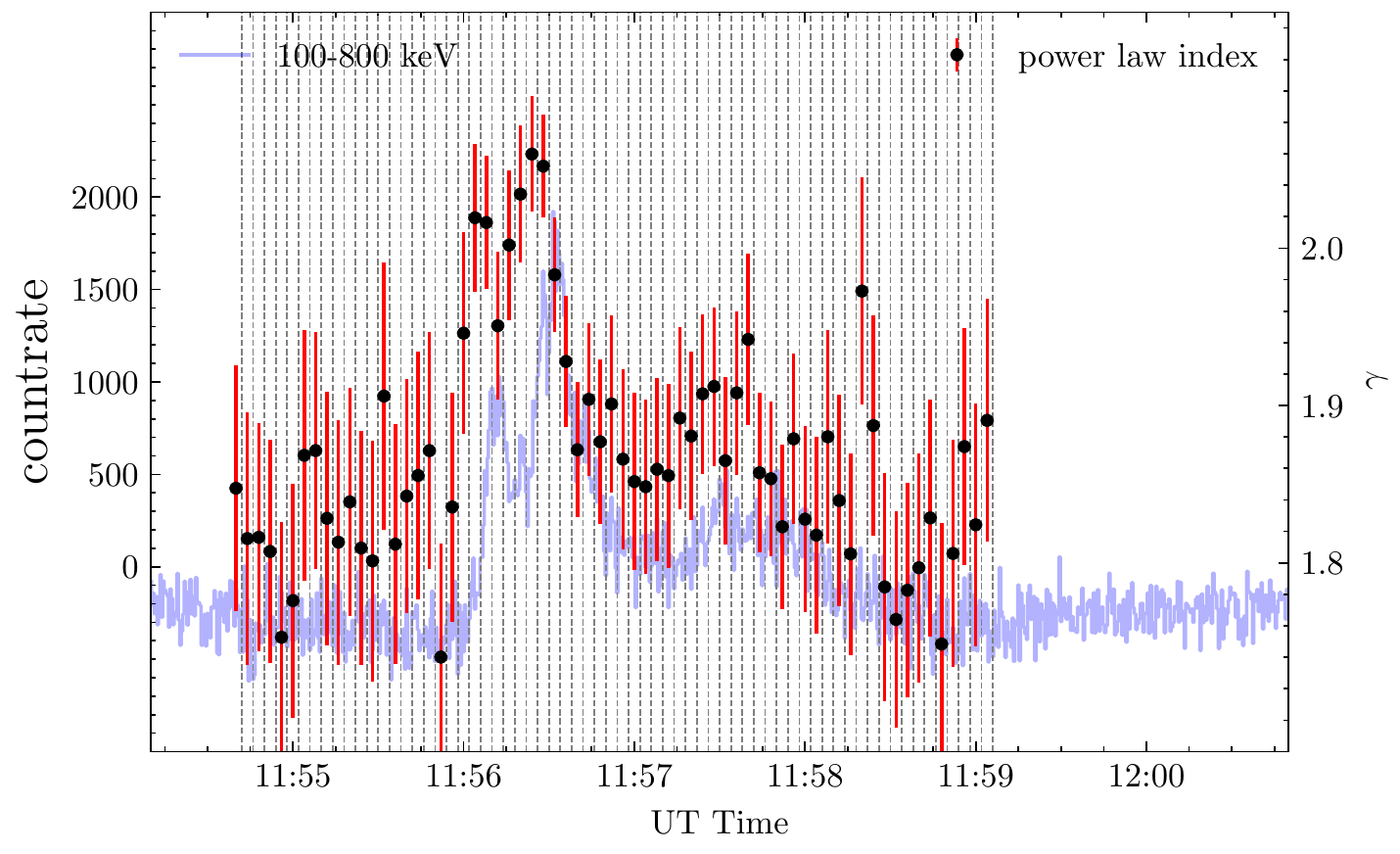}
\caption{The spectral power-law index evolution (by intervals) during the impulsive phase of flare, blue light curve is the 100-800 keV light curve , the black dot with red error bar is the power law index . \label{fig:hxmtfits}}
\end{figure}

 As shown in the figure \ref{fig:hxmtspec}, we derived the non-thermal electron index from HXMT observations, as initial electron beam constrains for the GS numerical model, and then estimated the transverse magnetic field and the non-thermal electrons density in the flaring GS source. To take the advantage of the continue HXMT/CsI observations, we made fitting to the CsI spectrum for all ROI intervals marked at figure \ref{fig:lc} with a simple power law model. In the figure \ref{fig:hxmtfits}, the power law photon index $\gamma$ follow the trail of light curve during the ROI time intervals, with a little difference from 1.7 to 2.1, which is consistent with the results present by \cite{2019ApJ...877..145L}. Even though \cite{2019ApJ...877..145L} shows a soft-hard-soft pattern from 11:56 UT to 11:57:20 UT at low-energy part and soft-hard-harder pattern at high-energy regime by KW data, but the HXMT/CsI spectral evolution of such narrow continuum bremsstrahlung tended to show a complex pattern almost identical to the light-curve pattern from 11:55 UT to 11:59 UT.

 Due to the time limit of the KW spectral observation, one could not get the picture of spectral evolution during the whole flaring phase. The soft-hard-soft pattern could be also seen from 11:56 UT to 11:57:20 UT in HXMT/CsI observation. For a longer time scale from 11:55 UT to 11:59 UT, the HXMT/CsI observations suggested a hard-soft-hard spectral pattern. In general, flaring is dominated by energetic electrons, and the study presented by \cite{2019ApJ...877..145L} implies that only very weak ion acceleration contributes to the hard X-ray continuum emission, supporting a prolonged non-thermal electron acceleration during the main phase. Repeated magnetic reconnections release bulk of energy, consequently accelerate the electrons. Energetic non-thermal electrons penetrated to the footpoints from corona loop top, emitting the hard X-ray photons repeatedly, which also verified the correlated QPPs of hard X-ray and 2.7 GHz emission.

\begin{figure}[hbt!]
\centering
\plotone{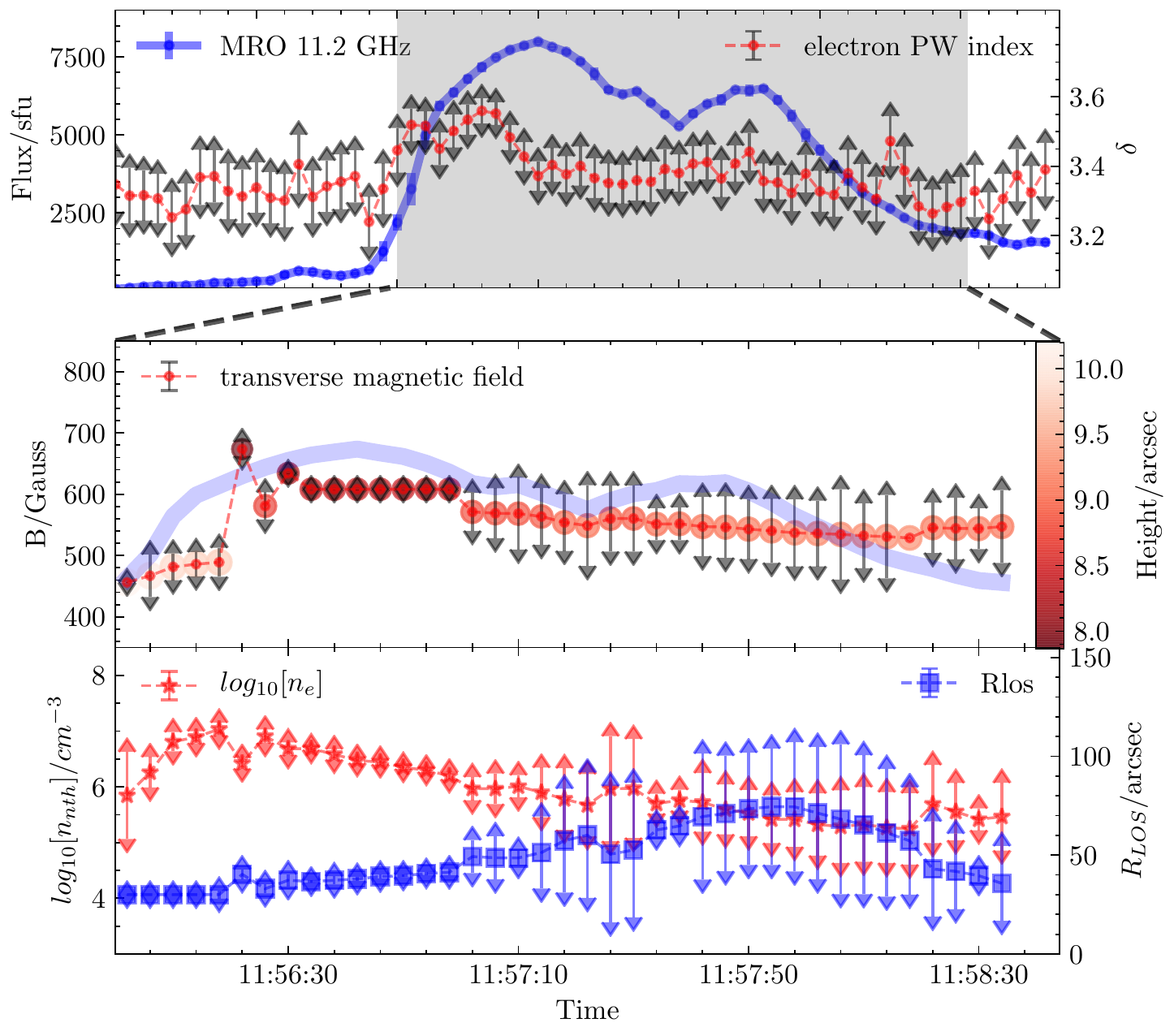}
\caption{The radio spectral fitting results during the impulsive phase of flare: in the top panel light blue curve is the MRO 11.2 GHz light curve, red dot with error bar is the inferred electron power law index $\gamma_{e}$; in the middle panel we shown the mean transverse magnetic field evolution, the transparent blue curve is the MRO 11.2 GHz light-curve; in the bottom panel red star with error bar is the logarithm non thermal electron density, blue square with error bar is the line of sight radio emission source size respectively from the spectral fittings. \label{fig:gyrospec}}
\end{figure}

To illustrate the evolution of non-thermal electron acceleration during the X9.3 flare, we modeled the GS source with the same method for all gyrosynchrotron microwave spectral intervals, so that we derived the evolution of the mean corona magnetic field and non-thermal electrons. As shown in figure \ref{fig:gyrospec}, the mean transverse magnetic field slowly increased at the impulsive phase, and reaches $\sim$ 700 Guass near the hard X-ray peak, then slightly decayed during the main flaring phase, almost had the same trend as the MRO 111.2 GHz light-curve. Furthermore, the non-thermal electron density also tends to have a similar behavior. However on the contrary we could see the line of sight radio emitting source size grew during the whole impulsive phase. For typical impulsive flares, the GS microwave spectrum peak frequency is around 10 GHz, corresponding to a $\sim$ 600 Gauss magnetic field \citep{1982ApJ...259..350D,1985ARA&A..23..169D}, which is consistent with the values in our fittings for the X9.3 flare.

The bottom panel in the figure \ref{fig:gyrospec} shows non-thermal electron density  and the GS source beam radius evolution from GS modeling. We could see that the non-thermal electrons are well correlated with MRO 11.2GHz light-curve, non-thermal electron beams accelerated and penetrated into deeper corona before the main HXMT hard X-ray phase. Later the non-thermal electrons lose their energy in the 'thick target' immediately, non-thermal electrons density decreases, heating the base corona or even the chromosphere, which might be an explanation to the mid-IR emission discussed in \cite{2018SpWea..16.1261G}. However during the whole flaring process, GS source beam size and the mean transverse magnetic field did not show drastic changes (see the results present in the figure \ref{fig:gyrospec}), those findings also support our hypothesis of the GS source and its initial configuration.

Meanwhile if we consider a rough magnetic field model in the solar atmosphere $B(h)=0.5(h/R_{\sun})^{-1.5}$, we could derive the height above photosphere which varies from 7.8 $\arcsec$ to 10.0 $\arcsec$ (about 5.7 Mm to 7.4 Mm), as shown in the second panel in figure \ref{fig:gyrospec}. The height evolution suggests a lower corona GS source. On the other hand, we could see a persistent expansion of the background hot plasma source during the acceleration phase, and decrease along GS emission while the energy dissipation, implying the bulk of magnetic reconnection energy dissipate by accelerated non-thermal electrons and background plasma heating.

Based on the discussion in the previous sections and the GS source model fitting results, theoretical time profile of RSTN observations could be estimated from the gyrosynchrotron approximation. Then in this study , the gyrosynchrotron approximation equation 2.15 describe in \cite{2011SSRv..159..225W} could be rewritten as following:

\begin{equation}
S_{s f u}=2.1 \times 10^{-28.0-2.83 \delta} f_{\mathrm{GHz}}^{1.20-0.90 \delta} B^{-0.20+0.90 \delta}\int \mathscr{N}_{\text {e}}LdA
\label{eq:flux}
\end{equation}

\begin{figure}[hbt!]
\centering
\plotone{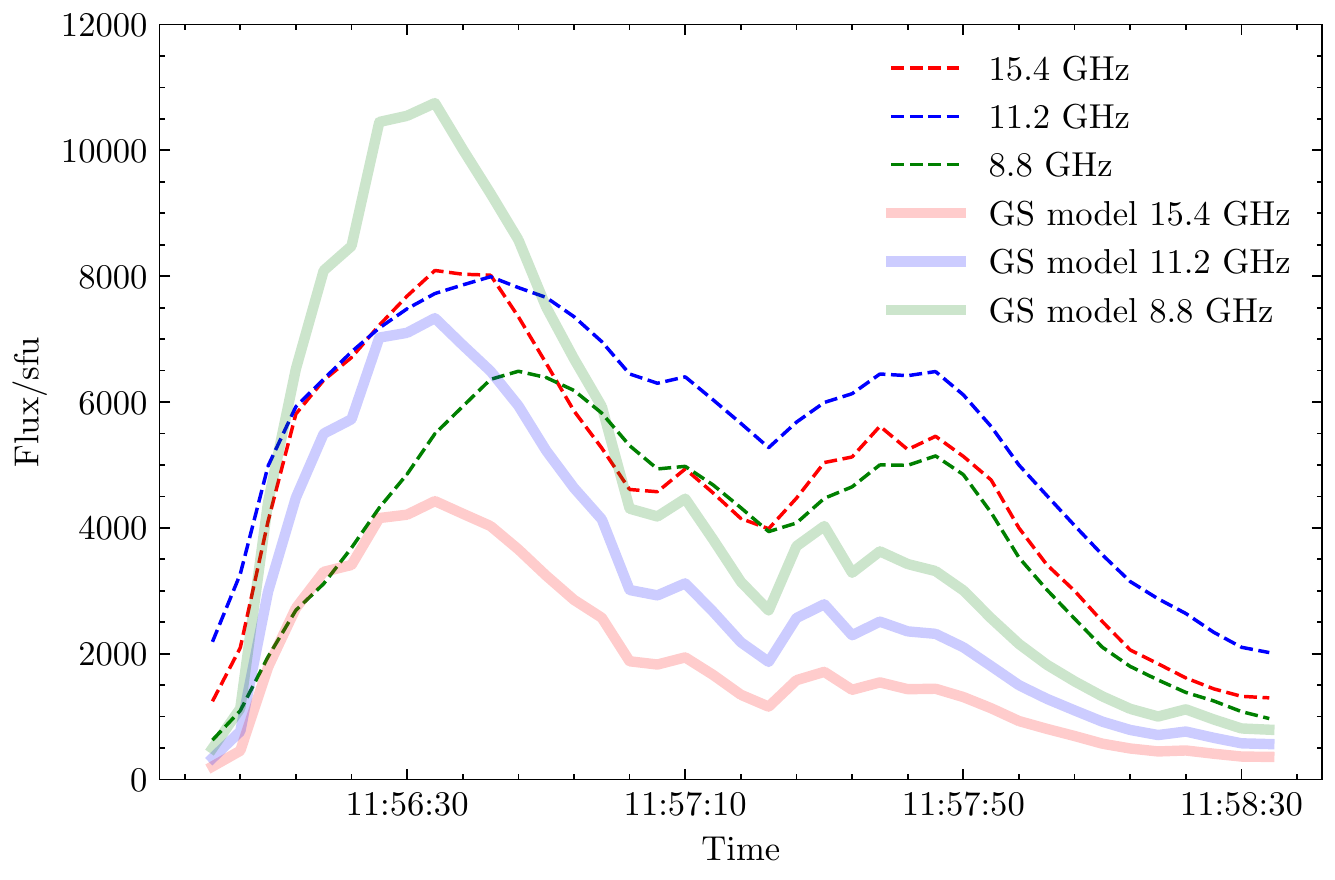}
\caption{Comparison between the GS modeling radio light-curve and observational light-curve for 8.8 GHz, 11.2 GHz and 15.4 GHz. The red, blue and green dashed lines stand for observational data respectively, the widened lines are 15.4 GHz ,11.2 GHz and 8.8 GHz modeling results respectively. \label{fig:gyrosfu}}
\end{figure}

The total number of non-thermal electrons is $\mathscr{N}_{\text {tot}} = \int \mathscr{N}_{\text {e}}LdA $, where L is the column depth, $\mathscr{N}_{\text {e}}$ is the non-thermal electron density. However in a homogeneous cylinder model, $\mathscr{N}_{\text {\rm tot}} =\pi \mathscr{N}_{\text {e}}L R_{\rm los}^2 $, where $R_{\rm los}$ is the GS source beam radius estimated from GS modeling. Then we could derived the modeling flux from equation \ref{eq:flux} (see figure \ref{fig:gyrosfu}). During the main impulsive phase the modeling flux is well correlated to the observed one near 11.2 GHz, but in the decay phase, the correlation is not good, which may be due to the complexity of GS source evolution in the active region.

\section{Conclusion and Summary}

By using the HXMT CsI hard X-ray data, we conduct a comprehensive timing and spectral analysis of the largest flare during solar cycle 24 with GHz radio data. The correlation of hard X-ray observations and radio observations reveal that both emissions might origin from common energetic electron source. Base on such hypothesis, we built up a simple homogeneous cylinder GS source model, utilized both hard X-ray and radio spectra, and figured out the evolution of GS source during the X9.3 flare.

The HXMT/CsI data indicated a power law distribution for the hard X-ray continue photon spectrum, which originated from a softer power law distribution non-thermal electrons. However, for the same population of non-thermal electrons, in higher magnetic field strengths, the production efficiency for different wave bands is rather similar, but for a weak magnetic field, hard X-ray is much efficient than radio by lower-energy electrons\citep{2020ApJ...894..158K}. During the impulsive phase, the gyrosynchrotron microwave spectral fittings strongly indicated that magnetic reconnection released bulk of energy which can accelerate a large population of non-thermal electrons, and those energetic electrons are accelerated to the dense corona base, losing their energy immediately and heating the background plasma at the same time. Indeed, those non-thermal electrons accelerated by huge magnetic energy, will produce rich radio emission through the magnetized plasma in the corona, and emitting hard X-ray photons by bremsstrahlung process.

As shown in figure \ref{fig:gyrosfu}, the theoretical radio flux is mismatched with observations with few times difference. The time delay between observations and GS modeling indicated that the hard X-ray source is different from the GS source, and energetic electron beams may take few seconds to travel. Low energy non-thermal electrons easily lose their energy through the magnetized plasma and even are trapped in the loops, but for high energy electrons they could be injected deeper to the lower corona. Those low energy electrons also could be accelerated repeatedly by repeated energy release from magnetic reconnection point, which might explain the common QPP behaviours of hard X-rays and 2.7 GHz. However the QPP at $\sim 22 $ s is almost 3 times bigger than the time lag, probably because energetic electrons acceleration time scale is much more shorter than repeated magnetic energy release during the impulsive phase or the non-thermal electrons are  accelerated from a lower corona source. And the QPP at $\sim $ 80 s supports a long live GS source over the whole flare. Even though the estimated mean transverse corona magnetic field also demonstrated a low corona origin source, we still have to realize the complexity of flaring process.


\acknowledgments

This work is supported by the National Natural Science Foundation of China (Grants No. U1838103, 11622326, U1838201, U1838202, U1631242, 11820101002), the National Program on Key Research and Development Project (Grants No. 2016YFA0400803, 2016YFA0400800). We thank Dr.Juha Kallunki from Metsähovi Radio Observatory for data calibration.

%

\vspace{5mm}
\facilities{HXMT(CsI), RSTN, GOES, SDO(AIA and HMI), Metsähovi Radio Observatory}


\software{\textbf{XSPEC\citep{1996ASPC..101...17A},
HEAsoft\citep{2014ascl.soft08004N},
astropy\citep{2013A&A...558A..33A},
SolarSoft\citep{2012ascl.soft08013F}
}}



\begin{figure}[ht!]
\epsscale{0.55}
\plotone{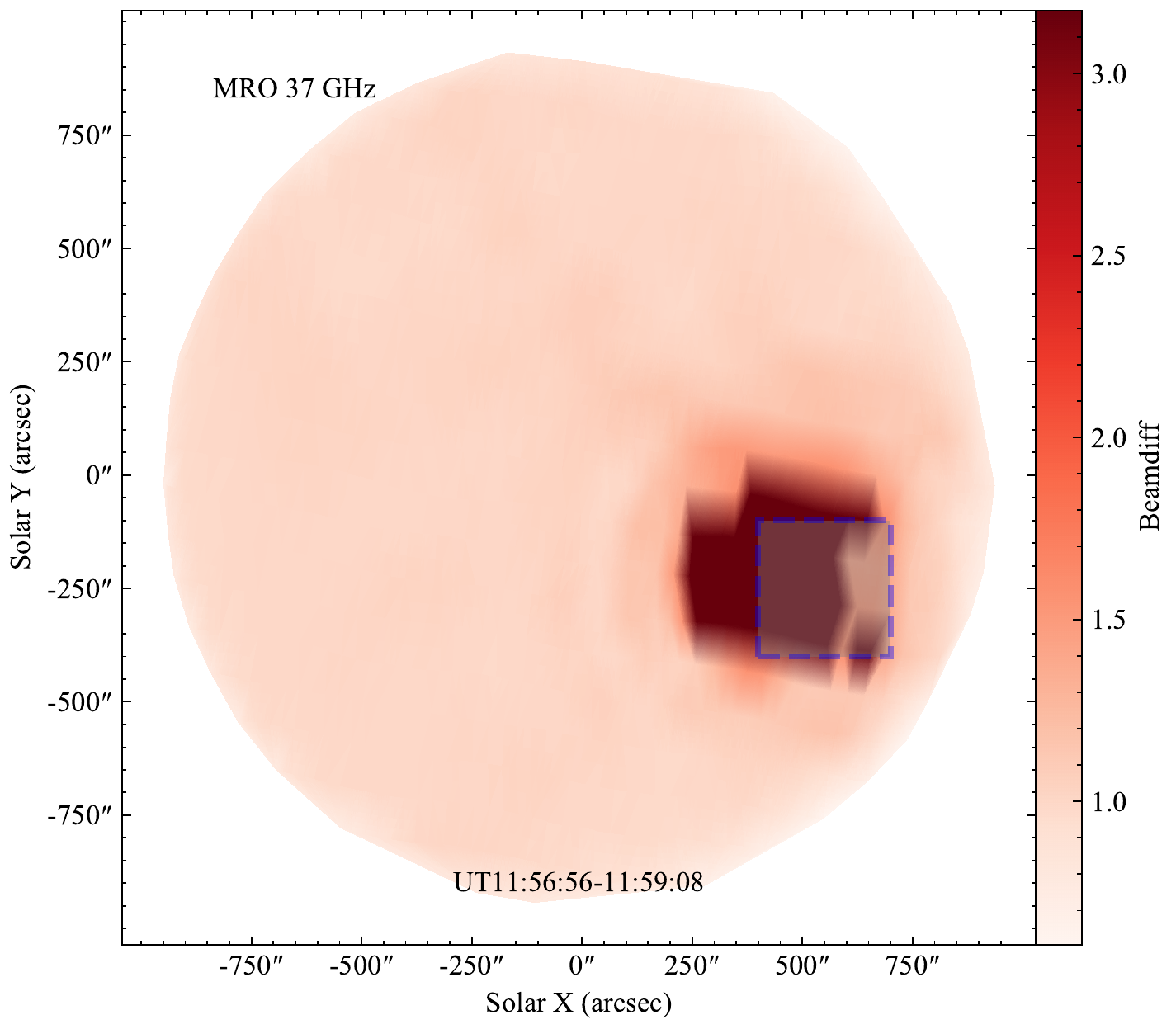}
\caption{The figure shows the MRO 37 GHz radio map from 11:56:56 to 11:59:08 UT, the dashed gray rectangle mark the ROI region of the X9.3 flare. \label{fig:mromap}}
\end{figure}

\appendix
Figure \ref{fig:mromap} shows the gyrosynchrotron source observed by Metsähovi Radio Observatory from 11:56:56 to 11:59:08 UT.


\bibliography{20170906hxmt}{}
\bibliographystyle{aasjournal}



\end{document}